\documentstyle{article}
\begin{document}
\title{\bf \Large Searching a systematics for nonfactorizable
contributions to hadronic decays of $ D^{0}$ and $ D^{+}$ mesons}
\vskip 0.5 cm
\author { {\bf R. C. Verma}\\
\normalsize Centre for Advanced Study in Physics, Department of Physics,\\
\normalsize Panjab University, Chandigarh -160014 {\bf India} }
\vskip 0.5 cm
\maketitle
\begin{abstract}
\large  \baselineskip 24pt We investigate nonfactorizable contributions to
charm meson
decays in $ D \rightarrow \bar K \pi $/ $ \bar K \rho$ /$ \bar K ^{*}
\pi$ / $ \bar K a_{1}$ / $ \bar K ^{*} \rho $ modes. Obtaining the
contributions from spectator-quark diagrams for $ N_{c}$ = 3, we
determine nonfactorizable isospin 1/2 and 3/2 amplitudes required to
explain the data for these modes. We observe that ratio of these
amplitudes seem to follow a universal value.
\end{abstract}
\newpage\large \section{Introduction} \par \baselineskip 24pt  As
precise data [1] on some of the weak hadronic and semileptonic decays of
charm mesons is available, it has now become possible to test the
validity of the factorization model. Employing isospin formalism, in a
strong interaction -phase independent manner, recently, Kamal and Pham
[2] has, shown that the naive factorization model fails to account for
isospin amplitudes ( $ A_{1/2}$ and $ A_{3/2}$ ) for  $ D \rightarrow
\bar K \pi$, $D \rightarrow \bar K \rho $, and $ D \rightarrow \bar
K^{*} \pi$ modes. It was observed that for $ D \rightarrow $ PP (P $
\equiv $ pseudoscalar meson), whereas the factorization assumption
accounts for branching ratio of $ D ^{+} \rightarrow \bar K^{0}
\pi^{+}$, it overestimates $ A_{1/2} $ for $ D^{0}$ -decays. For $ D
\rightarrow PV$ ( $ V \equiv $ vector mesons) modes, it overestimates $
A_{3/2} $ for $ D \rightarrow \bar K \rho $, decays and underestimates $
A_{3/2} $ for $ D \rightarrow K^{*} \pi$ decays. One of the ways to
remove the discrepancy could be the inelastic final state interaction
[2,3] which can feed $ \bar K^{*} \pi $ channel at the expense of $ \bar
K^{*} \rho $ channel. An alternative may be the nonfactorizable
contributions [4] arising through the soft gluon exchange, which are
generally ignored in the factorization model. Recently, there has been a
growing interest [5,6] in exploring such contributions in the hadronic
decays of charmed and bottom mesons. Major cause of the concern has been
that $ N_{c} \rightarrow \infty $ limit, which is considered to be
supported by D-meson phenomenology, fails when carried over to B-meson
decays [7]. Therefore, a reinvestigation of the charm decays is called
for. \par \baselineskip 24pt We, in this paper, study the
nonfactorizable contributions to $ D \rightarrow \bar K \pi $, $ D
\rightarrow \bar K \rho $, $ D \rightarrow \bar K ^ {*} \pi$, $D
\rightarrow \bar K a_{1} $ and $ D \rightarrow \bar K^ {*} \rho $
decays. Employing the isospin formalism in the phase-independent manner,
we determine these contributions in respective 1/2 and 3/2 -isospin
amplitudes from experiment and search for a systematics in the
amplitudes.\\ \par  \baselineskip 24pt In section 2, weak Hamiltonian is
discussed. In next sections, we study these decay modes one by one.
Summary and discussion are given in the last section.
 \large
\section{Weak Hamiltonian} \par \baselineskip 24pt   The effective weak
Hamiltonian for Cabibbo-angle-enhanced decays of the charm hadrons is
given by $$ H_{w} \hskip 0.5 cm  = \hskip 0.5 cm \tilde{G}_{F} [c_{1} (
\bar u d)( \bar s c) + c_{2} ( \bar s d) ( \bar u c) ],
\eqno(1)$$\baselineskip 24pt  where $ \tilde{G}_{F} = \frac {G_{F}} {
\sqrt2} V_{ud} V_{cs}^{*} $ and $ \bar q_{1} q_{2}$ represents color
singlet V - A current $$ \bar q_{1} q_{2} \equiv \bar q_{1} \gamma_{\mu}
(1 - \gamma_{5} ) q_{2},$$\baselineskip 24pt    and the QCD coefficients
at the charm mass scale are [8]  $$   c_{1} = 1.26 \pm 0.04, \hskip 2.5
cm c_{2} = - 0.51 \pm 0.05. \eqno(2)$$  \baselineskip 24pt  Due to the
Fierz transformation of the product of two Dirac currents in (1) in $
N_{c}$-color space, the Hamiltonian takes the following form $$ H_{w}
^{CF} \hskip 0.5 cm = \hskip 0.5 cm\tilde{G}_{F} [a_{1} ( \bar u d)(
\bar s c) + c_{2} H_{w}^{8} ],$$ $$ H_{w} ^{CS} \hskip 0.5 cm = \hskip
0.5 cm \tilde{G}_{F} [a_{2} ( \bar s d ) ( \bar u c ) + c_{1}
\tilde{H}_{w}^{8} ], \eqno(3)$$ \baselineskip 24pt   for color favored
(CF) and color suppressed (CS) decay amplitudes, and $$ a_{1,2} =
c_{1,2} + \frac {c_{2,1}} {N_{c}},$$ $$ H^{8}_{w} \hskip 0.5 cm = \hskip
0.5 cm \frac {1} {2} \sum_{a=1}^{8} ( \bar u \lambda ^{a} d ) ( \bar s
\lambda ^{a} c ),$$ $$ \tilde{H} ^ {8}_ {w} \hskip 0.5 cm = \hskip 0.5
cm \frac {1} {2} \sum_{a=1}^{8} ( \bar s \lambda ^{a} d ) ( \bar u
\lambda ^{a} c ), \eqno(4)$$ \baselineskip 24pt   where color-octet
currents $$ \bar q_{1} \lambda ^{a} q_{2} \hskip 0.5 cm \equiv \hskip
0.5 cm \bar q _{1} \gamma_{\mu} (1 - \gamma_{5} ) \lambda ^{a} q_{2} $$
\baselineskip 24pt involve the Gell-Mann matrices, $ \lambda ^{a} $ for
color. Matrix element of the first terms of (3) can be calculated using
the factorization scheme [9]. In this scheme, nonfactorizable
contributions from the second terms in (3) are ignored, so long one
restricts to color singlet intermediate states. Due to the neglect of
these terms, one usually treats $ a_{1}$ and $ a_{2}$ as input
parameters instead of using $ N_{c} = 3$ in reality. Empirically $ D
\rightarrow \bar K \pi$ data seem to favor $ N_{c} \rightarrow \infty $
limit [9] which is justified with the hope that the nonperturbative
effects arising due to the soft gluon exchange between the colored-octet
current in (4), relative to the factorizable amplitudes arising from the
color singlet current in (3), are of the order $ \frac {1} {N_{c}}$ in
the large $ N_{c}$-limit. In fact, in some of the QCD calculations [10],
it has been claimed that for $ B \rightarrow D \pi $, and $ D
\rightarrow \bar K \pi $ as well, nonfactorizable terms tend to cancel
the contributions from $ \frac {1} { N_{c}}$ terms in the first terms of
(3). However, B-mesons don't favor this result empirically. Further,
this does not guarantee that such cancellations would persist for other
decay modes too. In other words, $ a_{1}$ and $ a_{2}$ are not universal
parameters, these are decay dependent if one is to stick to the
factorization model. An alternative is to take $ N_{c} = 3$ and
investigate nonfactorizable contributions more seriously. This is what
we do in the following sections employing the flavor-isospin framework.
\section {$ D \rightarrow \bar K \pi$ decays} \par In terms of the
isospin amplitudes and final state interactions phases [2], $$ A(D^{0}
\rightarrow K^{-} \pi^{+} ) \hskip 0.5 cm = \hskip 0.5 cm \frac {1} {
\sqrt3} [ A_{3/2} e^{i \delta_{3/2}} + \sqrt2 A_{1/2} e^{i \delta_{1/2}}
], $$ $$ A(D^{0} \rightarrow \bar K^{0} \pi^{0} ) \hskip 0.5 cm = \hskip
0.5 cm \frac {1} { \sqrt3} [ \sqrt2 A_{3/2} e^{i \delta_{3/2}} - A_{1/2}
e^{i \delta_{1/2}} ], $$ $$ A(D^{+} \rightarrow \bar K^{0} \pi^{+} )
\hskip 0.5 cm = \hskip 0.5 cm \sqrt3 A_{3/2} e^{i \delta_{3/2}}.
\eqno(5) $$ \baselineskip 24pt   These lead to the following phase
independent quantities: $$ | A(D^{0} \rightarrow K^{-} \pi^{+} ) |^{2}
+ | A ( D^{0} \rightarrow \bar K^{0} \pi^{0} )|^{2} \hskip 0.5 cm =
\hskip 0.5 cm | A_{1/2} |^{2} + | A_{3/2} |^{2},$$ $$ | A(D^{+}
\rightarrow \bar K^{0} \pi^{+} ) |^{2} \hskip 0.5 cm = \hskip 0.5 cm  3
| A_{3/2} |^{2}. \eqno(6)$$ \baselineskip 24pt   With Experimental [1]
values: $$ \hskip 1 cm B ( D^{0} \rightarrow K^{-} \pi^{+} ) \hskip 1 cm
= \hskip 1 cm (4.01 \pm 0.14) \% ,$$ $$ \hskip 1cm B ( D^{0} \rightarrow
\bar K^{0} \pi^{0} ) \hskip 1 cm = \hskip 1 cm (2.05 \pm 0.26) \%,$$ $$
\hskip 1 cm B(D^{+} \rightarrow \bar K^{0} \pi^{+} )  \hskip 1 cm =
\hskip 1 cm (2.74 \pm 0.29) \%,$$ \baselineskip 24pt and D-meson
lifetimes $$ \tau_{D^{0}} \hskip 0.5 cm = \hskip 0.5 cm 0.415 \hskip 0.1
cm ps,$$ $$ \tau_{D^{+}} \hskip 0.5 cm = \hskip 0.5 cm 1.057 \hskip 0.1
cm ps,$$ \baselineskip 24pt   decay rate formula $$ \Gamma (D
\rightarrow PP ) \hskip 0.5 cm = \hskip 0.5 cm  | \tilde{G}_{F} |^{2}
\frac {p} {8 \pi m_{D}^{2}} | A ( D \rightarrow PP ) |^{2} $$ yields $$
\hskip 1 cm | A_{1/2} |_{exp} \hskip 1 cm =\hskip 1 cm 0.387 \pm 0.011
\hskip 0.1 cm GeV^{3},$$ $$ \hskip 1 cm |A_{3/2} |_{exp} \hskip 1 cm =
\hskip 1 cm 0.097 \pm 0.005 \hskip 0.1 cm GeV^{3} \eqno(7)$$
\baselineskip 24pt   Since the relations (6) are independent of the
final state interaction phases, one might well evaluate them without the
phases and determine nonfactorizable contributions using experimental
values. We separate the factorizable and nonfactorizable parts of the
decay amplitude as $$ A ( D \rightarrow \bar K \pi ) = A^{f} ( D
\rightarrow \bar K \pi ) +A^{nf} ( D \rightarrow \bar K \pi ).
\eqno(8)$$ \baselineskip 24pt   Using the factorization scheme,
factorizable part of the decay amplitudes can be written [9] as $$ A^{f}
( D^{0} \rightarrow K^{-} \pi^{+} ) \hskip 0.4 cm = \hskip 0.4 cm a_{1}
f_ \pi ( m_{D}^{2} - m_{K}^{2} ) F_{0}^{DK} (m_ \pi ^ {2} ),$$  $$
\hskip 0.2 cm = \hskip 0.2 cm 0.351 \hskip 0.1 cm GeV^{3},$$ $$ A^{f} (
D^{0} \rightarrow \bar K^{0} \pi^{0} ) \hskip 0.4 cm = \hskip 0.4 cm
\frac {1}{ \sqrt2} a_{2} f_ {K} ( m_{D}^{2} - m_{ \pi}^{2} ) F_{0}^{D
\pi} (m_{K}^ {2} ),$$ $$ \hskip 0.4 cm = \hskip 0.4 cm -0.030 \hskip 0.1
cm GeV^{3},$$ $$ A^{f} ( D^{+} \rightarrow \bar K^{0} \pi^{+} ) \hskip
0.4 cm = \hskip 0.4 cm a_{1} f_ \pi ( m_{D}^{2} - m_{K}^{2} ) F_{0}^{DK}
(m_ \pi ^ {2} ) $$ $$+ a_{2} f_ {K} ( m_{D}^{2} - m_{ \pi}^{2} )
F_{0}^{D \pi} (m_{K}^ {2} ),$$ $$ \hskip 0.4 cm = \hskip 0.4 cm 0.309
\hskip 0.1 cm GeV^{3} \eqno(9)$$ \baselineskip 24pt   Numerical input
for these terms is taken as $$ a_{1} = 1.09 , \hskip 2.8 cm a_{2} = -
0.09 ,$$ $$ f_{ \pi} = 0.132 \hskip 0.1 cm GeV, \hskip 2.8 cm f_{K} =
0.161 \hskip 0.1 cm GeV,$$ and $$ F_{0}^{DK} (0) = 0.76 , \hskip 2.8 cm
F_{0}^{D \pi} (0) = 0.83  \eqno(10)$$ \baselineskip 24pt from Ref [11].
Using isospin C. G. Coefficients, nonfactorizable part of the decay
amplitudes can be expressed as scattering amplitudes for spurion + $ D
\rightarrow \bar K + \pi$ process: $$ A^{nf} ( D^{0} \rightarrow K^{-}
\pi^{+}) \hskip 0.4 cm = \hskip 0.4 cm \frac {1}{3} c_{2} ( < \bar K
\pi||H_{w}^{8} || D >_{3/2} + 2 < \bar K \pi || H_{w}^{8} || D >_{1/2}
),$$ $$ A^{nf} ( D^{0} \rightarrow \bar K^{0} \pi^{0}) \hskip 0.4 cm =
\hskip 0.4 cm \frac { \sqrt2}{3} c_{1} ( < \bar K \pi||
\tilde{H}_{w}^{8} || D >_{3/2} - < \bar K \pi || \tilde{H}_{w}^{8} || D
>_{1/2} ),$$ $$ A^{nf} ( D^{+} \rightarrow \bar K^{0} \pi^{+}) \hskip
0.4 cm = \hskip 0.4 cm c_{2}  < \bar K \pi|| \tilde{H}_{w}^{8} || D
>_{3/2} + c_{1} < \bar K \pi || \bar H_{w}^{8} || D >_{3/2}  \eqno(11)$$
\baselineskip 24pt   In order to reduce the number of unknown reduced
amplitude, we assume the following $$ < \bar K \pi|| H_{w}^{8} || D
>_{1/2} \hskip 1 cm = \hskip 1 cm < \bar K \pi || \bar H_{w}^{8} || D
>_{1/2} , \eqno(12) $$ $$ < \bar K \pi|| H_{w}^{8} || D >_{3/2} \hskip 1
cm = \hskip 1 cm < \bar K \pi || \tilde{H}_{w}^{8} || D >_{3/2} ,
\eqno(13)$$ \baselineskip 24pt  as both $ H_{w}^{8}$ and $
\tilde{H}_{w}^{8}$ behave like $|1,1>$ component of an isovector
spurion. In fact,  $ H_{w}^{8}$ and $\tilde{H}_{w}^{8}$ transform into
each other under the interchange of u and s quarks. Thus, in the limit of
flavor
SU(3) symmetry, the constraints given in eqs. (12) and (13) become reliable.
{}From (5), we can write $$ A_{1/2}^{nf} ( D \rightarrow \bar K
\pi ) \hskip 0.2 cm = \hskip 0.2 cm \frac{1}{ \sqrt3} \{ \sqrt2 A^{nf}
(D^{0} \rightarrow K^{-} \pi^{+} ) - A^{nf} (D^{0} \rightarrow \bar
K^{0} \pi^{0}) \}$$ $$ A_{3/2}^{nf} ( D \rightarrow \bar K \pi ) \hskip
0.2 cm = \hskip 0.2 cm \frac{1}{ \sqrt3} \{ A^{nf} (D^{0} \rightarrow
K^{-} \pi^{+} ) + \sqrt2 A^{nf} (D^{0} \rightarrow \bar K^{0} \pi^{0})
\}$$ $$ \hskip 0.2 cm = \hskip 0.2 cm \frac{1}{ \sqrt3} \{ A^{nf} (D^{+}
\rightarrow \bar K^{0} \pi^{+} ) \} \eqno(14)$$ \baselineskip 24pt
Relations (12) and (13) then lead to the following prediction: $$ \frac
{ A_{1/2}^{nf} (D \rightarrow \bar K \pi )}{ A_{3/2}^{nf} (D \rightarrow
\bar K \pi )} \hskip 0.5 cm = \hskip 0.5 cm \frac {c_{1}^{2} + 2
c_{2}^{2}}{ \sqrt2 (c_{2}^{2} - c_{1}^{2})}$$ $$ \hskip 0.5 cm = \hskip
0.5 cm -1.123 \pm 0.112. \eqno(15)$$ \baselineskip 24pt Experimentally,
the nonfactorized isospin amplitudes is determined as $$ A_{1/2}^{nf}
\hskip 0.5 cm = \hskip 0.5 cm + 0.082 \pm 0.032 \hskip 0.1 cm GeV^{3},$$
$$ A_{3/2}^{nf} \hskip 0.5 cm = \hskip 0.5 cm - 0.081 \pm 0.023 \hskip
0.1 cm GeV^{3}, \eqno(16)$$ from $ A_{1/2} = 0.387 \pm 0.011 \hskip 0.1
cm GeV^{3}$, and $ A_{3/2} = 0.097 \pm 0.005 \hskip 0.1 cm GeV^{3}$
given in (7). This yields $$ \frac { A_{1/2}^{nf}}{A_{3/2}^{nf}} \hskip
0.5 cm = \hskip 0.5 cm - 1.011 \pm 0.250, \eqno(17)$$ \baselineskip 24pt
in good agreement with theoretical expectation (15). Such isospin
formalism can easily be extended to $ D \rightarrow \bar K \rho$, $ D
\rightarrow \bar K^{*} \pi$, and $ D \rightarrow \bar K a_{1}$ decays.
Since the isospin structure of these decay modes is exactly the same as
given in (5) and (11), the same value $-1.123  \pm 0.112$ would follow
for the respective ratio of the nonfactorizable isospin amplitudes $
A_{1/2}^{nf}$ and $ A_{3/2}^{nf}$ in each of these cases. In the
following, we determine these amplitudes from experimental values of the
branching ratios of these decay modes, and compare their ratio with
theoretically expected one. \section {  $ D \rightarrow \bar K \rho$
decays} \par \baselineskip 24pt We begin with the definition of the
decay amplitude $ A ( D \rightarrow PV ) $ through the decay rate
formula, $$ \Gamma ( D \rightarrow PV ) \hskip 0.5 cm = \hskip 0.5 cm |
\tilde{G}_{F} |^{2} \frac {p^{3}}{2 \pi} | A ( D \rightarrow PV )
|^{2}.$$ $ A (D \rightarrow PV ) $ has the dimension of $\hskip 0.1 cm
GeV$ and is obtained by writing down the standard decay amplitude and
removing from it a factor $ 2 m_{V} ( \epsilon^{*} . P_{D} )
\tilde{G}_{F}$, where $m_{V}$ is the vector meson mass and $
\epsilon^{*} $ is its polarization vector, $P_{D}$ is the D-meson four
momentum. Factorizable parts of the decay amplitudes $ A ( D \rightarrow
\bar K \rho ) $ can be written as $$ A^{f} ( D^{0} \rightarrow K^{-}
\rho^{+} ) \hskip 0.5 cm = \hskip 0.5 cm a_{1} f_ \rho F_{1}^{DK} (m_
\rho ^ {2} ),$$  $$ \hskip 0.5 cm = \hskip 0.5 cm 0.203 \hskip 0.1 cm
GeV,$$ $$ A^{f} ( D^{0} \rightarrow \bar K^{0} \rho^{0} ) \hskip 0.5 cm
= \hskip 0.5 cm \frac {1}{ \sqrt2} a_{2} f_ {K} A_{0}^{D \rho} (m_{K}^
{2} ),$$ $$ \hskip 0.5 cm = \hskip 0.5 cm -0.0074 \hskip 0.1 cm GeV,$$
$$ A^{f} ( D^{+} \rightarrow \bar K^{0} \rho^{+} ) \hskip 0.5 cm =
\hskip 0.5 cm a_{1} f_ \rho F_{1}^{DK} (m_ \rho ^ {2} ) + a_{2} f_ {K}
A_{0}^{D \rho} (m_{K}^ {2} ),$$  $$ \hskip 0.5 cm = \hskip 0.5 cm 0.192
\hskip 0.1 cm GeV.$$ Here, we use $$ f_{ \rho} \hskip 0.5 cm = \hskip
0.5 cm 0.212 \hskip 0.1 cm GeV,$$ $$ F_{1}^{DK} (0) \hskip 0.5 cm =
\hskip 0.5 cm F_{0}^{DK} (0) \hskip 0.5 cm = \hskip 0.5 cm 0.76,$$ from
(10) and  $$ A_{0}^{D \rho} (0) \hskip 0.5 cm = \hskip 0.5 cm 0.669$$
\baselineskip 24pt    is taken from the BSW model [9]. Numerical values
given above are calculated by extrapolating $ F_{1}^{DK} (q^{2})$ and $
A_{0}^{D \rho}(q^{2})$ using a monopole formulae with pole mass $ 2.11
\hskip 0.1 cm GeV$ ($ D_{s}^{*}$ pole) and $1.865 \hskip 0.1 cm GeV $ (
D-pole) respectivelly. Experimental [1] values of branching ratios: $$
\hskip 1 cm B ( D^{0} \rightarrow K^{-} \rho^{+} ) \hskip 1 cm = \hskip
1 cm (10.4 \pm 1.3) \% ,$$ $$ \hskip 1cm B ( D^{0} \rightarrow \bar
K^{0} \rho^{0} ) \hskip 1 cm = \hskip 1 cm (1.10 \pm 0.18) \%,$$ $$
\hskip 1 cm  B(D^{+} \rightarrow \bar K^{0} \rho^{+} )  \hskip 1 cm =
\hskip 1 cm  (6.6 \pm 2.5) \%,$$ \baselineskip 24pt yield the total
isospin-amplitudes: $$ \hskip 1 cm | A_{1/2} |_{exp} \hskip 1 cm =\hskip
1 cm 0.235 \pm 0.014 \hskip 0.1 cm GeV,$$ $$ \hskip 1 cm |A_{3/2}
|_{exp} \hskip 1 cm = \hskip 1 cm 0.067 \pm 0.013 \hskip 0.1 cm GeV
\eqno(19)$$ \baselineskip 24pt Writing $ A^{nf} ( D \rightarrow \bar K
\rho )$ analogues to (11) - (13), inturn leads to $$ A_{1/2}^{nf} (D
\rightarrow \bar K \rho)  \hskip 0.5 cm = \hskip 0.5 cm + 0.065 \pm
0.015 \hskip 0.1 cm GeV,$$ $$ A_{3/2}^{nf} ( D \rightarrow \bar K \rho)
\hskip 0.5 cm = \hskip 0.5 cm - 0.041 \pm 0.013 \hskip 0.1 cm GeV,
\eqno(20)$$ \baselineskip 24pt   with positive signs chosen for both $
A_{1/2} $ and $ A_{3/2}$ in (19). The ratio $$ \frac { A_{1/2}^{nf} (D
\rightarrow \bar K \rho )}{A_{3/2}^{nf} (D \rightarrow \bar K \rho )}
\hskip 0.5 cm = \hskip 0.5 cm -1.481 \pm 0.582, \eqno(21)$$
\baselineskip 24pt is consistent with (15) within errors. \section { $ D
\rightarrow \bar K^{*} \pi$ decays} \par   Repeating the same procedure
used for $ D \rightarrow \bar K \rho $ decays, here the factorizable
amplitudes are given by $$ A^{f} ( D^{0} \rightarrow \bar K^{* -}
\pi^{+} ) \hskip 0.5 cm = \hskip 0.5 cm a_{1} f_ \pi A_{0}^{DK^{*}} (m_
\pi ^ {2} ),$$  $$ \hskip 0.5 cm = \hskip 0.5 cm 0.100 \hskip 0.1 cm
GeV,$$ $$ A^{f} ( D^{0} \rightarrow \bar K^{* 0} \pi^{0} ) \hskip 0.5 cm
= \hskip 0.5 cm \frac {1}{ \sqrt2} a_{2} f_ {K^{*}} F_{1}^{D \pi}
(m_{K^{*}}^ {2} ),$$ $$ \hskip 0.5 cm = \hskip 0.5 cm -0.015 \hskip 0.1
cm GeV,$$ $$ A^{f} ( D^{+} \rightarrow \bar K^{* 0} \pi^{+} ) \hskip 0.5
cm = \hskip 0.5 cm a_{1} f_ \pi A_{0}^{DK^{*}} (m_ \pi ^ {2} ) + a_{2}
f_ {K^{*}} F_{1}^{D \pi} (m_{K^{*}}^ {2} ),$$  $$ \hskip 0.5 cm = \hskip
0.5 cm 0.080 \hskip 0.1 cm GeV. \eqno(22)$$ \baselineskip 24pt Here, we
use $$ F_{1}^{D \pi} (0) \hskip 0.5 cm = \hskip 0.5 cm F_{0}^{D \pi} (0)
\hskip 0.5 cm = \hskip 0.5 cm 0.83, \hskip 0.5 cm f_{K^{*}} \hskip 0.5
cm = \hskip 0.5 cm 0.221 \hskip 0.1 cm GeV,$$ and $ A_{0}^{DK^{*}} (0) $
is determined from the relation $$ A_{0}^{DK^{*}}(0) = A_{3}^{DK^{*}}(0)
= \frac {1}{2m_{K^{*}}} \{ (m_{D} + m_{K^{*}})
A_{1}^{DK^{*}}(0)-(m_{D}-m_{K^{*}})A_{2}^{DK^{*}}(0) \}. \eqno(23)$$
\baselineskip 24pt Semileptonic $ D \rightarrow K^{*}e \nu $ data [11]
yields $$ A_{1}^{DK^{*}}(0) \hskip 0.5 cm = \hskip 0.5 cm 0.61 \pm 0.05,
\hskip 1 cm A_{2}^{DK^{*}}(0) \hskip 0.5 cm = \hskip 0.5 cm 0.45 \pm
0.09. $$ Which gives $$ A_{0}^{DK^{*}}(0) \hskip 0.5 cm = \hskip 0.5 cm
0.70 \pm 0.09.$$   Experimental [1] values of branching ratios: $$
\hskip 1 cm B ( D^{0} \rightarrow K^{* -} \pi^{+} ) \hskip 1 cm = \hskip
1 cm (4.9 \pm 0.6) \% ,$$ $$ \hskip 1cm B ( D^{0} \rightarrow \bar K^{*
0} \pi^{0} ) \hskip 1 cm = \hskip 1 cm (3.0 \pm 0.4) \%,$$ $$ \hskip 1
cm B(D^{+} \rightarrow \bar K^{* 0} \pi^{+} ) \hskip 1 cm = \hskip 1 cm
(2.2 \pm 0.4) \%, \eqno(24)$$\baselineskip 24pt  yield the total
isospin-amplitude: $$ \hskip 1 cm | A_{1/2} |_{exp} \hskip 1 cm =\hskip
1 cm 0.186 \pm 0.009 \hskip 0.1 cm GeV,$$ $$ \hskip 1 cm |A_{3/2}
|_{exp} \hskip 1 cm = \hskip 1 cm -0.036 \pm 0.003 \hskip 0.1 cm GeV
\eqno(25)$$ \baselineskip 24pt   Choosing positive and negative sign for
the $ A_{1/2}$ and $ A_{3/2}$ terms respectively, we find $$
A_{1/2}^{nf} (D \rightarrow \bar K^{*} \pi) \hskip 0.5 cm = \hskip 0.5
cm + 0.096 \pm 0.009 \hskip 0.1 cm GeV,$$ $$ A_{3/2}^{nf} ( D
\rightarrow \bar K^{*} \pi ) \hskip 0.5 cm = \hskip 0.5 cm - 0.082 \pm
0.008 \hskip 0.1 cm GeV, \eqno(26)$$ \baselineskip 24pt leading to $$
\frac { A_{1/2}^{nf} (D \rightarrow \bar K^{*} \pi )}{A_{3/2}^{nf} (D
\rightarrow \bar K^{*} \pi )} \hskip 0.5 cm = \hskip 0.5 cm  -1.171 \pm
0.158, \eqno(27)$$ \baselineskip 24pt for monopole like extrapolation of
$ F_{1}^{D \pi}(q^{2})$. This ratio is consistent with theoretical
expectation (15) within errors. Also note that $$ |A_{1/2}^{nf}(D
\rightarrow K^{*} \pi)| > |A_{1/2}^{nf}(D \rightarrow \bar K \rho ) |,
\eqno(28)$$   \baselineskip 24pt consistent with theoretical
expectations that final state having low momentum is likely to be
affected more by the soft gluon exchange effects. \section{  $ D
\rightarrow \bar K a_{1}$ decays} \par  \baselineskip 24pt  $ D
\rightarrow \bar K a_{1}$ decays can also be treated in a manner similar
to that used for $ D \rightarrow \bar K \rho $ modes; due to the
similarity in their Lorentz structure. For $ D \rightarrow \bar K a_{1}$
decays, factorized amplitudes are ( upto the scale factor $ \frac {
\tilde{G}_{F}}{ \sqrt2} ( \epsilon^{*}.p)2m_{a_{1}}$ ): $$ A^{f} ( D^{0}
\rightarrow K^{-} a_{1}^{+} ) \hskip 0.5 cm = \hskip 0.5 cm a_{1} f_{
a_{1}} F_{1}^{DK} (m_{a_{1}} ^ {2} ),$$  $$ \hskip 0.5 cm = \hskip 0.5
cm 0.285 \hskip 0.1 cm GeV,$$ $$ A^{f} ( D^{0} \rightarrow \bar K^{0}
a_{1}^{0} ) \hskip 0.5 cm = \hskip 0.5 cm \frac {1}{ \sqrt2} a_{2} f_
{K} V_{0}^{D a_{1}} (m_{K}^ {2} ),$$ $$ \hskip 0.5 cm = \hskip 0.5 cm
0,$$ $$ A^{f} ( D^{+} \rightarrow \bar K^{0} a_{1}^{+} ) \hskip 0.5 cm =
\hskip 0.5 cm a_{1} f_{a_{1}} F_{1}^{DK} (m_{a_{1}}^{2} ) + a_{2} f_ {K}
V_{0}^{D a_{1}} (m_{K}^ {2} ),$$  $$ \hskip 0.5 cm = \hskip 0.5 cm 0.285
\hskip 0.1 cm GeV. \eqno(29) $$ Here, we use $$ f_{ a_{1}} \hskip 0.5 cm
= \hskip 0.5 cm 0.221 \hskip 0.1 cm GeV,$$\baselineskip 24pt  and take
$$ V_{0}^{DK} (0) \hskip 0.5 cm \approx \hskip 0.5 cm 0$$ \baselineskip
24pt due to the orthogonality of the D and $ a_{1}$ spin-wave functions.
The experimental values [1,12] for branching ratios: $$ \hskip 1 cm B (
D^{0} \rightarrow K^{-} a_{1}^{+} ) \hskip 1 cm = \hskip 1 cm (7.9 \pm
1.2) \% ,$$ $$ \hskip 1cm B ( D^{0} \rightarrow \bar K^{0} a_{1}^{0} )
\hskip 1 cm = \hskip 1 cm (0.43 \pm 0.99) \%,$$ $$ \hskip 1 cm  B(D^{+}
\rightarrow \bar K^{0} a_{1}^{+} ) \hskip 1 cm = \hskip 1 cm  (8.1 \pm
1.7) \%, \eqno(30)$$ \baselineskip 24pt   when used for analogues of
relations (6), yield total isospin amplitudes: $$ \hskip 1 cm | A_{1/2}
|_{exp} \hskip 1 cm =\hskip 1 cm (0.582_{-0.055}^{+0.066}) \hskip 0.1 cm
GeV,$$ $$ \hskip 1 cm |A_{3/2} |_{exp} \hskip 1 cm = \hskip 1 cm
(0.338_{-0.064}^{+0.077}) \hskip 0.1 cm GeV. \eqno(31)$$ \baselineskip
24pt Here we have neglected the small effects arising due to the width
of $ a_{1}$ meson [13]. Choosing positive and negative signs for $
A_{1/2}$ and $ A_{3/2}$ respectively in (31), we find  $$ A_{1/2}^{nf}
(D \rightarrow \bar K a_{1}) \hskip 0.5 cm = \hskip 0.5 cm + 0.349 \pm
0.060 \hskip 0.1 cm GeV,$$ $$ A_{3/2}^{nf} ( D \rightarrow \bar K a_{1})
\hskip 0.5 cm = \hskip 0.5 cm - 0.221 \pm 0.023 \hskip 0.1 cm GeV,
\eqno(32)$$ leading to $$ \frac { A_{1/2}^{nf} (D \rightarrow \bar K
a_{1} )}{A_{3/2}^{nf} (D \rightarrow \bar K a_{1} )} \hskip 0.5 cm =
\hskip 0.5 cm -0.910 \pm 0.165, \eqno(33)$$ \baselineskip 24pt which is
consistent with theoretical expectation (15). Also note that $$
|A_{1/2}^{nf}(D \rightarrow PA)| > |A_{1/2}^{nf}(D \rightarrow PV ) |,
\eqno(34)$$ \baselineskip 24pt again in accordance with theoretical
expectation, as in the $ D \rightarrow \bar K a_{1}$ decays, final state
momentum is smaller than that in $ D \rightarrow PV $ mode.  \par
\baselineskip 24pt   Thus, we observe that in all the decay modes,
considered so far, $ D \rightarrow \bar K \pi $, $ D \rightarrow \bar K
\rho$, $ D \rightarrow \bar K^{*} \pi $, $ D \rightarrow \bar K a_{1}$,
the nonfactorizable isospin amplitude $ A_{1/2}^{nf}$ not only have the
same sign for these decays, but also bears the same ratio, with in the
experimental errors, with $A_{3/2}^{nf}$ amplitude, i.e., $$ \frac {
A_{1/2}^{nf} (D \rightarrow \bar K a_{1} )}{A_{3/2}^{nf} (D \rightarrow
\bar K a_{1} )} \hskip 0.5 cm \approx \hskip 0.5 cm \frac { A_{1/2}^{nf}
(D \rightarrow \bar K^{*} \pi )}{A_{3/2}^{nf} (D \rightarrow \bar K^{*}
\pi )} \hskip 0.5 cm \approx \hskip 0.5 cm\frac { A_{1/2}^{nf} (D
\rightarrow \bar K \rho )}{A_{3/2}^{nf} (D \rightarrow \bar K \rho )}$$
$$\hskip 0.5 cm \approx \hskip 0.5 cm \frac { A_{1/2}^{nf} (D
\rightarrow \bar K \pi )}{A_{3/2}^{nf} (D \rightarrow \bar K \pi )}.
\eqno(35) $$ \baselineskip 24pt   Further, we notice that the
nonfactorized amplitudes show an increasing pattern with decreasing
momenta available to the final state particles, i.e., $$ |A^{nf} (D
\rightarrow \bar K a_{1} )| \hskip 0.5 cm > \hskip 0.5 cm |A^{nf} (D
\rightarrow \bar K^{*} \pi )| \hskip 0.5 cm > \hskip 0.5 cm | A^{nf} (D
\rightarrow \bar K \rho)| \eqno(36)$$\baselineskip 24pt  This behaviour
is understandable, since low momentum states are likely to be affected
more through the exchange of soft gluons and can acquire larger
nonfactorizable contributions. If one takes value of the ratio of $
A_{1/2}^{nf}$ and $ A_{3/2}^{nf}$ in (35) to be -1.123 as obtained in
(15), and determine $ A_{3/2}^{nf}$ from $D^{+}$ decay, one can predict
the sum of the branching ratios of $D^{0}$ -meson decays in the
corresponding mode. Following this procedure, we predict  $$ B(D^{0}
\rightarrow K^{-} \pi^{+}) + B(D^{0} \rightarrow \bar K^{0} \pi^{0})
\hskip 0.3 cm = 6.30 \pm 0.67 \% $$ $$ = ( 6.06 \pm 0.30 \% \hskip 0.5
cm Expt),$$ $$ B(D^{0} \rightarrow K^{-} \rho^{+}) + B(D^{0} \rightarrow
\bar K^{0} \rho^{0}) \hskip 0.4 cm = 10.17 \pm 3.85 \% $$ $$ =(11.50 \pm
1.31 \% \hskip 0.5 cm Expt ) ,$$ $$ B(D^{0} \rightarrow \bar K^{*-}
\pi^{+}) + B(D^{0} \rightarrow \bar K^{*0} \pi^{0}) = 6.29 \pm 1.20 \%$$
$$ =(7.9 \pm 2.2 \%, \hskip 0.5cm Expt)$$ $$ B(D^{0} \rightarrow K^{-}
a_{1}^{+}) + B(D^{0} \rightarrow \bar K^{0} a_{1}^{0}) \hskip 0.4 cm =
10.67 \pm 2.24 \% $$ $$ =(8.33 \pm 1.56 \% \hskip 0.5 cm Expt).
\eqno(37)$$ \baselineskip 24pt    All theoretical values match well with
experiment. Infact, these relations can be expressed in a general form
as $$ B_{-+} + B_{00} = \frac { \tau_{D^{0}}}{3 \tau_{D^{+}}} B_{0+} [ 1
+ \{ \alpha + \frac {( \sqrt2 - \alpha ) A_{-+}^{fac} - (1+ \sqrt2
\alpha)A_{00}^{fac}}{A_{0+}} \}^{2}], \eqno(38)$$ \baselineskip 24pt
with $ \alpha \equiv A_{1/2}^{nf}/A_{3/2}^{nf},$ where subscript $ -+$,
$00$, $0+$ denote the charge states of strange and nonstrange mesons
emitted in each case. $ A_{-+}^{fac}$ and $ A_{00}^{fac}$ denote the
factorized amplitudes of $ D^{0}$ decays. $ A_{0+}$ is obtained from the
$D^{+}$ -decay branching ratio $ B_{0+}$ , $$ A_{0+} \hskip 0.5 cm =
\hskip 0.5 cm \sqrt{ \frac { B_{0+}}{ \tau_{D^{+}} \times (Kinematic
factors)} }$$ \section { $ D \rightarrow \bar K^{*} \rho $ decays} \par
\baselineskip 24pt  In general, $ D \rightarrow VV $ modes involve
Lorentz structure for three partial waves: S, P, D waves. Therefore, one
may expect nonfactorizable contributions to be present in all of them.
Experimentally their partial wave structure has been analysed [1,12].
Data indicates that S-wave is dominant in the $ D^{+} \rightarrow \bar
K^{*0} \rho^{+} $ decays. For $ D^{0} \rightarrow \bar K^{*0} \rho^{0}$
mode, D-wave component seems to exist which interfere destructively with
S-wave. P-wave component of these two modes is negligible. However, for
$ D^{0} \rightarrow K^{-} \rho^{+}$, data is not clean enough to
separate these partial waves, though P-wave component is found to be
small here also. Looking at the experimental status, we introduce the
nonfactorizable term in S-wave, and relate only the S-wave decay
branching ratios. Then $D \rightarrow VV$ decays also share the same
isospin structure as given in (5) and (11). The factorizable decay
amplitudes ( in S-wave ) are given by (upto the scale factor $
\tilde{G}_{F}  \epsilon_{1}^{*} . \epsilon_{2}^{*} $ ) $$ A^{f} ( D^{0}
\rightarrow \bar K^{*-} \rho^{+} ) \hskip 0.5 cm = \hskip 0.5 cm a_{1} (
m_{D} + m_{K^{*}}) m_{ \rho} f_ \rho A_{1}^{DK^{*}} (m_ \rho ^ {2} ),$$
$$ \hskip 0.5 cm = \hskip 0.5 cm 0.330 \hskip 0.1 cm GeV^{3},$$ $$ A^{f}
( D^{0} \rightarrow \bar K^{* 0} \rho^{0} ) \hskip 0.5 cm = \hskip 0.5
cm \frac {1}{ \sqrt2} a_{2} (m_{D}+m_{ \rho}) f_ {K^{*}} m_{K^{*}}
A_{1}^{D \rho} (m_{K^{*}}^ {2} ),$$ $$ \hskip 0.5 cm = \hskip 0.5 cm
-0.030 \hskip 0.1 cm GeV^{3},$$ $$ A^{f} ( D^{+} \rightarrow \bar K^{*
0} \rho^{+} ) \hskip 0.5 cm = \hskip 0.5 cm a_{1} (m_{D}+m_{K^{*}})f_
\rho m_{K^{*}} A_{1}^{DK^{*}} (m_ \rho ^ {2} )$$ $$ \hskip 0.5 cm +
\hskip 0.5 cm a_{2} (m_{D}+m_{ \rho})f_ {K{*}} A_{1}^{D \rho}
(m_{K^{*}}^ {2} ),$$  $$ \hskip 0.5 cm = \hskip 0.5 cm 0.289 \hskip 0.1
cm GeV. \eqno(39)$$ \baselineskip 24pt   Numerical values given above
are determined using the form factors $ A_{1}^{DK^{*}}(0)$ already given
and $ A_{1}^{D \rho}(0) = 0.775$ as determined in the BSW model [9].
\par The decay rate formula [14] reduces to $$ \Gamma (D \rightarrow VV
) = | \tilde{G}_{F} |^{2} \frac {p} {8 \pi m_{D}^{2}} ( 2 + ( \frac
{m_{D}^{2}-m_{V_{1}}^{2}-m_{V_{2}}^{2}}{2m_{V_{1}}m_{V_{2}}})^2) | A ( D
\rightarrow VV ) |^{2}, \eqno(40) $$ \baselineskip 24pt  for S-wave.
Writing $$ A(D \rightarrow VV) \hskip 0.5 cm = \hskip 0.5 cm A^{f} (D
\rightarrow VV) + A^{nf} (D \rightarrow VV),$$ we obtain $$
A_{3/2}^{nf}(D \rightarrow VV) \hskip 0.5 cm = \hskip 0.5 cm
+0.114_{-0.040}^{+0.021}, \eqno(41)$$ \baselineskip 24pt from
experimental value of S-wave branching ratio $ B_{S} ( D^{+} \rightarrow
\bar K^{*0} \rho^{+}) = (1.7 \pm 1.6) \%.$ Extending the apparent
universality (35) of the ratio to the $ D \rightarrow VV$ decay modes,
$$ \frac {A_{1/2}^{nf} (D\rightarrow VV)}{A_{3/2}^{nf} ( D \rightarrow
VV )} \hskip 0.5 cm = \hskip 0.5 cm - 1.123 \pm 0.112.$$  \baselineskip
24pt  We calculate the sum of S-wave branching ratios of $ D^{0}
\rightarrow K^{*-} \rho^{+}$ and $ D^{0} \rightarrow \bar K^{*0}
\rho^{0}$ decays, $$ B_{S}(D^{0} \rightarrow K^{*-} \rho^{+}) +
B_{S}(D^{0} \rightarrow \bar K^{*0} \rho^{0}) \hskip 0.5 cm = \hskip 0.5
cm ( 14.0_{-1.3}^{+2.9}) \%. \eqno(42)$$ Subtracting the experimentally
known value of $$ B_{S}(D^{0} \rightarrow \bar K^{*0} \rho^{0}) \hskip
0.5 cm = \hskip 0.5 cm (3.0 \pm 0.6) \%,$$ we predict $$ B_{S}(D^{0}
\rightarrow K^{*-} \rho^{0}) \hskip 0.5 cm = \hskip 0.5 cm
(11.0_{-1.9}^{+3.5}) \%.$$  \baselineskip 24pt  It is interesting to
remark this value satisfy the following relation: $$ \frac
{B_{total}(D^{0} \rightarrow K^{*-} \rho^{+})}{B_{S}(D^{0} \rightarrow
K^{*-} \rho^{+})} \hskip 0.5 cm = \hskip 0.5 cm \frac {B_{total}(D^{0}
\rightarrow \bar K^{*0} \rho^{0})}{B_{S}(D^{0} \rightarrow \bar K^{*0}
\rho^{0})},$$ $$ 0.54 \pm 0.25 \hskip 0.5 cm = \hskip 0.5 cm 0.53 \pm
0.17,$$ indicating the destructive interference of S- and D-wave partial
waves for $ D^{0} \rightarrow \bar K^{*-} \rho^{+}$ decay also. \large
\section { Summary and Discussion} \par  \baselineskip 24pt  In this
work, we have investigated the nonfactorizable contributions to various
decays of $ D^{0}$ and $D^{+}$ mesons  $ \bar K \pi / \bar K \rho / \bar
K^{*} \pi / \bar K a_{1} / \bar K^{*} \rho$ states involving isospin 1/2
and 3/2 final states. In our analysis, we take the real value of $ N_{c}
= 3$. Since the nonfactorizable contributions, being nonperturbative,
are difficult to be calculated, we determine their amplitudes in these
isospin states required by the experiment. We have ignored the
annihilation contributions here as these don't contribute to $ D^{+}$
decays and for $D^{0}$ -decays these are suppressed due to the small
value of $a_{2} ( \approx -0.09 \pm 0.05)$. We observe that not only the
nonfactorizable isospin amplitudes $ A_{1/2}^{nf}$ have the sign for the
modes considered, but also bears the same ratio with $ A_{3/2}^{nf}$
within experimental errors. The ratio can be understood on the basis of
V-spin symmetry, under which nonfactorizable parts $ H_{w}^{8}$ and $
\tilde{H}_{w}^{8}$ of the weak Hamiltonian transform into each other.
Further the nonfactorizable contributions also show an increasing
pattern with decreasing momentum available to the final state particles
emitted in these decays. Extending the apparent universality of the
ratio $ A_{1/2}^{nf} / A_{3/2}^{nf}$ to $ D \rightarrow VV$ modes, we
predict the S-wave branching ratio for $ B_{S}(D^{0} \rightarrow K^{*-}
\rho^{+}) = (11.0_{-1.9}^{+3.5}) \%$, indicating destructive
interference between S-wave and D-wave components for this decay.

 \vskip 3 cm

 \maketitle
  \par  RCV thanks A.N. Kamal for providing
 support from a grant from NSERC, Canada.  He also thanks and the
 Theoretical Physics, Institute, Department of Physics, University of
 Alberta, where most of the work was done, for their hospitality.

 \newpage
 \maketitle
 
\end{document}